\documentclass[12pt]{amsart}
\usepackage{amssymb}
\usepackage{amsmath,amsthm,amsfonts}
\usepackage[applemac]{inputenc}
\usepackage{graphicx}
\usepackage{bbm}
\usepackage{pdfsync}
\usepackage{fancyhdr}
\usepackage{mathrsfs}
\usepackage{slashed}
\usepackage{color}
\usepackage{xcolor}
\usepackage[breaklinks,colorlinks,backref]{hyperref}

\hypersetup{
colorlinks, 
linktoc=all, 
linkcolor=blue, 
citecolor=hyptxt,
urlcolor=blue}
\hypersetup{
citebordercolor=,
filebordercolor=green,
linkbordercolor=blue
}
\definecolor{hyptxt}{rgb}{0.7, 0.4, 0.9}

\usepackage{bibtopic}

\usepackage[full]{textcomp} 
\usepackage{fbb} 
\usepackage[varqu,varl]{inconsolata}
\usepackage[libertine,bigdelims,vvarbb]{newtxmath}
\usepackage[cal=boondoxo]{mathalfa}
\usepackage[T1]{fontenc}
\useosf

\newtheorem{prop}{Proposition}[section]

\newcommand{\beprop}{\begin{prop}}
\newcommand{\enprop}{\end{prop}}
\newcommand{\bprf}{\begin{proof}}
\newcommand{\eprf}{\end{proof}}

\newcommand{\ket}[1]{|\kern.3ex#1\kern.3ex\rangle}
\newcommand{\bra}[1]{\langle\kern.3ex #1 \kern.3ex|}
\newcommand{\scalar}[2]{\langle\kern.3ex #1 \kern.3ex|\kern.3ex#2\kern.3ex\rangle}

\def\R{\mathbb{R}}

\def\C{\mathbb{C}}

\definecolor{hervecolor}{rgb}{0.8,0,0.7}

\numberwithin{equation}{section}

\def\R{{\rm I\hspace{-.15em}R}}

\def\1{\mbox{I\hspace{-.15em}1}}

\def\b{\begin{equation}}
\def\e{\end{equation}}

\begin{document}
\date{\today}
\title{Conceptual and technical challenges\\ of quantum gravity}
\author{M.V. Takook}

\address{\emph{ APC, UMR 7164}\\
\emph{Universit\'e de Paris} \\
\emph{75205 Paris, France}}

\email{ takook@apc.in2p3.fr}

{\abstract{The appearance of infinity together with collapsing quantum state due to the observation or
interaction, which are two challenging features of quantum field theory, become very serious
problems in quantum gravity as well as in quantum geometry of space-time. These problems can be divided into two categories: technical and conceptual parts. In the technical aspect, the biggest problem comes from the definition of the point, which constructs classical geometry. By changing the point definition, the technical problem may be approximately bypassed, and a mathematical formulation of quantum geometry may be found. On the other hand, the conceptual problem comes from the quantum state collapse due to the observation since the entanglement between the observer and gravity cannot be eliminated in the presence of gravity. The conceptual part does not depend on the techniques which are employed. As much as possible, we try to discuss these problems by rediscovering the perception and understanding of a phenomenon for an observer in quantum theory.
}}

\maketitle

{\it Proposed PACS numbers}: 04.62.+v, 98.80.Cq, 12.10.Dm

\tableofcontents
\section{Introduction}

Quantum gravity is one of the most mysterious problems in theoretical physics, which has been considered for many years \cite{dois,is02,is04,buis,hpf,ag}. This article intends to present an overview of this problem differently. The appearance of infinity in the probability amplitude and the collapse of the quantum state due to an observation or interaction are two critical challenges of quantum field theory, which have been elegantly treated in quantum field theory in Minkowski space-time and have resulted in quantum engineering technology \cite{takook18}. The first problem, as mentioned above, has been solved by the regularization and renormalization technique \cite{weinb,ilio}. The second can be overcome using quantum states' superposition, and entanglement properties \cite{takook18}. These two challenges have also fatally appeared in quantum gravity. Therefore, the problems of quantum gravity or quantum geometry of space-time can also be divided into technical (appearance of infinity) and conceptual parts (entanglement and quantum state collapse).

In the technical part, the biggest problem comes from the point definition where the classical geometry is constructed. The geometry of space-time is a set of infinite points. The point definition is purely mathematical and does not have any dimension. It is also equivalent to nothing! Physically this definition is quite vague. By changing the point definition, the technical problem may be approximately bypassed, and a mathematical formulation of quantum geometry may be formulated. Care must be taken into account that quantum geometry is not discrete geometry. This one is a particular and exceptional case of quantization. The discussion of operators and the superposition of quantum states are our views on quantum geometry, which goes directly into the entanglement of different fields and geometry.

The conceptual part comes from the entanglement and collapsing quantum states and does not depend on the techniques used. This article discusses the conceptual problems of quantum gravity by redefining the perception and understanding of a phenomenon in quantum theory. We show that using the quantum theory principles, quantum gravity cannot lead us to find the quantum state of the universe, $\vert U \rangle$ (contain the observers) since the initial state is obscure. The approximation solution can only be discussed, which means the universe must be divided into two-part, the observer and the rest of the universe. Then the observer's rest of the universe quantum state, $\vert u,t \rangle$, can be discussed. Time $t$ is defined by the observer, and it is the time of observation. The best theoretical example of this approximate method to understand the quantum gravity effect is a black hole. There is enormous progress in understanding the quantum black hole in de Sitter space-time, which can be formulated in ambient space formalism \cite{takdsbl}. Recently an experiment was proposed which could confirm the quantum gravitational effect in the approximate method \cite{qgdetec}. They used the background field method with the Galilean background for space and time and Newtonian gravity for the gravitational field.

Unlike particles or waves (classical fields), which are appropriate notions for physical systems in the classical model, in the quantum theory model, a quantum state $\vert \alpha\rangle$ is a genuinely fundamental entity in the universe. In the quantum field model, the truth of a phenomenon is the quantum state, $\vert \alpha\rangle$, and the reality of a phenomenon is the collapsing of this quantum state, {\it i.e.} $\vert \beta \rangle$. The fact of a phenomenon (quality of observations) depends on: the observers and their measuring apparatus, the source producing the quantum initial state, and its time evolution. $\vert \alpha,t \rangle$ can be written in terms of the different bases of the Hilbert or Fock spaces, in which there is a unitary transformation between them. Although the choice of bases is arbitrary due to these unitary transformations, similar to the coordinate system selection in general relativity, each basis presents some properties of the quantum state or the possible reality observations. For instance, the number operator bases give the particle properties of the quantum state, although the superposition of the eigenstate of the number operator basis, such as classical wave, coherent states, squeezed state, etc., shows other properties of the quantum state.

Our goal here is to consider the problems of quantum gravity within the framework of the principles of quantum theory. In the presence of a gravitational field, since the concept of time depends on the observer, the initial state and its time evolution are the most challenging problems in quantum gravity. The observer and its measuring apparatus are entangled with the gravitational field. Then due to the collapsing of the quantum state, it is impossible to obtain or define the initial state explicitly, although it can be determined approximately. Therefore quantum gravity may be roughly understood by an observer, which is a part of the source of the gravitational field.

This article first recalls the basic concepts and principles of geometry, gravity, and quantum field theory (QFT). Then we apply these concepts and principles to consider quantum gravity. An approximation method, {\it i.e.} QFT in the de Sitter universe, will be discussed to understand the quantum gravity effects better. Finally, the perception and understanding of a phenomenon for an observer in quantum theory are redefined.

\section{Geometry and Gravity}

Geometry is explicitly defined by the metric tensor field, $g_{\mu\nu}(x)$, and the connection coefficients, $\Gamma_{\mu\nu}^{\rho}$, which is not a tensor field \cite{mtw73,naka}. The first is defined at a point, and the second is the onnection between a point with the neighboring points. Equivalently the geometry can be reformulated by the physical quantities of Riemann curvature tensor field, $R_{\mu\nu\rho\sigma}(x)$, torsion tensor field, $T_{\;\;\mu\nu}^{\rho}(x)$, and metric tensor field. It is important to note that in the general case, the geometry cannot be described by merely an irreducible rank-2 symmetric tensor field $g_{\mu\nu}$, and instead, the physical triplet $R_{\mu\nu\rho\sigma}, T_{\;\;\mu\nu}^{\rho}$ and $g_{\mu\nu}$ can be completely determined the properties of the geometry \cite{hehl,vsl10}. A special and important geometry is Weyl geometry where $\bigtriangledown_\mu g_{\rho\nu}=A_\mu g_{\rho\nu}$, which is invariant under the conformal transformation \cite{di73,fulton}. $A_\mu$ is a one-form field. By imposing the metric compatibility condition, $\bigtriangledown_\mu g_{\rho\nu}=0$, and torsion-free condition $\Gamma_{\mu\nu}^{\rho}=\Gamma_{\nu \mu}^{\rho}$, the geometry is defined by the Riemannian curvature tensor or equivalently by the metric tensor, which is called the Riemannian geometry.

From geometry, we know that a curved space-time manifold can be considered as a four-dimensional hyper-surface immersed in a flat space with dimensions larger than four. This hyper-surface is unique, but various coordinate systems can be selected, equivalent to different metrics. On the other hand, the different metrics are identical to the diverse observers.

Based on Einstein's general relativity, the gravitational field is equivalent to the Riemannian space-time geometry, and it is described by a fundamental rank-2 symmetric tensor field $g_{\mu\nu}$. The matter-radiation fields are immersed in space-time geometry and can also be considered a hyper-surface in space-time geometry. All of the matter-radiation fields are the source of the gravitational field or space-time hyper-surface, $M_4$, where their relations are defined by the Einstein field equation \cite{caro}:
\begin{equation} \label{eeq}
R_{\mu\nu}-\frac{1}{2}Rg_{\mu\nu}+\Lambda g_{\mu\nu}=8\pi GT_{\mu\nu}.
\end{equation}
$T_{\mu\nu}$ is the stress-energy tensor of all matter-radiation fields, $R_{\mu\nu}$ is the Ricci tensor, $R$ is the scalar curvature, $G$ is the gravitational constant, and $\Lambda$ is the cosmological constant.

Since the gravitational waves propagate on the light cone, we expect the field equations to be invariant under the conformal transformation. Still, Einstein's field equation is not invariant under the conformal transformation. It is a severe problem of Einstein's general relativity. For preserving the conformal invariant, in the linear approximation, the gravitational waves must be described by a rank-3 mix-symmetry tensor field, which can be conformal invariant and propagate on the light cone \cite{bfh83,tatafa10,tapeta12,amta19,rata19}.

In quantum theory, the gravitational field cannot be considered entirely as the space-time geometry since the quantum state of a particle in a gravitational field depends on the mass of the particle. However, classically due to the gravitational time dilation, at least one part of the gravity must appear as the space-time geometry \cite{mtw73}. Therefore, in quantum theory, the gravitational field must contain two elements, geometrical and non-geometrical. The second part is similar to the other matter-radiation fields. The quantum gravity returns to the quantization of these two parts. The geometry part of gravity is equivalent to space-time geometry, then the four-dimensional space-time manifold or hyper-surface $M_4$ must be quantized. In quantum geometry, matter-radiation fields and curved geometry are complexity entangled, which will be discussed in the section \ref{qg}.

\section{Quantum field theory}

From the classical point of view, a physical system or a phenomenon consists of particles and fields. From the quantum mechanical model, the quantum states, $\vert \alpha,t \rangle$, are the fundamental entities in the universe, which explains the physical system. In quantum theory, the classical quantities are represented by a Hermitian operator, and a quantum state defines the behavior of the physical system, which satisfies the following equation of motion in Minkowski space-time:
\begin{equation} \label{shoreq} i\hbar \frac{\partial}{\partial t}\vert\alpha,t\rangle=H\vert\alpha,t\rangle,\;\; \mbox{or} \;\;\vert\alpha,t\rangle=U(t,t_0;H)\vert\alpha,t_0\rangle, \end{equation}
where $H$ is the Hamiltonian of the system, $U(t,t_0;H)$ is the time evolution operator, and $|\alpha,t_0\rangle$ is the initial state.

In the first quantization, $H$ is the Hamiltonian of the particles, and the quantum state is immersed in the Hilbert space of the physical system, {\it i.e.} $\vert\alpha,t\rangle \in \mathcal{H}$. The Hilbert space can also be obtained from the physical system's operator algebra and symmetrical group. Hilbert space in quantum mechanics plays a similar role as space-time manifold, and their bases play the task of the coordinate system in classical mechanics. The operator is defined as a map from Hilbert space to Hilbert space (Operator: $\mathcal{H} \longrightarrow \mathcal{H}$).

In the second quantization or QFT, $H$ is the Hamiltonian of the radiation-matter fields, and the quantum state is immersed in the Fock space of the physical system, $\vert\alpha,t\rangle \in \mathcal{F}$. Fock space is constructed from the one-particle Hilbert space as:
$$ \mathcal{F}(\mathcal{H})=\left\{ \C, \mathcal{H}^{(1)}, \mathcal{H}^{(2)},\cdots, \mathcal{H}^{(n)}, \cdots \right\},$$
where $\C$ is vacuum state, $\mathcal{H}^{(1)}$ is one-particle states and $\mathcal{H}^{(n)}$ is n-particles states. n-particles states are constructed by tensor product of one-particle states (for bosons a symmetry product, $ \mathcal{H}^{(2)}=S\mathcal{H}^{(1)}\otimes \mathcal{H}^{(1)}$ and for fermions an anti-symmetric product, $ \mathcal{H}^{(2)}=A\mathcal{H}^{(1)}\otimes \mathcal{H}^{(1)}$).
The field operator is defined as a map from Fock space to Fock space:
$$ \mbox{Field operators}:\;\;\mathcal{F} \longrightarrow \mathcal{F}\, .$$

The equation of motion of the classical free field is a second-order differential equation, then it has two solutions, and each combination of them is also a solution. The selection of a solution is equivalent to choosing a vacuum state. The positive energy condition fixes a solution, and the vacuum state is selected. It is due to the existence of the time-like Killing vector field in Minkowski space ($H\propto \frac{\partial}{\partial t}$). It is a challenge for QFT in curved space-time where the time-like Killing vector field does not exist; one cannot define the vacuum state explicitly. Also, the time evolution of the quantum state is meaningless since the definition of time is observer-dependent, which will be discussed in the next section.

The first challenge of QFT is that the probability amplitude becomes infinite in the loop expansion. There exists two types of divergence: infrared divergences ($x-x'\longrightarrow \infty$) and ultraviolet divergence ($x-x'\longrightarrow 0$). Generally, the infinity appears in Feynman Green's functions $G_F(x-x')$. This infinity appears due to the definition of a point (zero) in space-time (ultraviolet divergence) or energy-momentum space (infrared divergences). This problem can be solved elegantly using standard renormalization methods in flat Minkowski space-time. The infinity problem in QFT is one of the technical challenges in quantum gravity.

From the principles of quantum theory, one cannot calculate and comprehend the quantum state $|\alpha,t\rangle$ because we need the initial state $|\alpha,t_0\rangle$, which must be obtained from the observation. We also need the observance of the quantum state for perception, but after it, due to collapsing of the quantum state $|\alpha,t_0\rangle$, we get a new quantum state $|\beta,t_0\rangle$: 
$$ |\alpha,t_0\rangle+\; \mbox{observation} \Longrightarrow |\beta,t_0\rangle \,. $$
Note that
$t_0$ is the time of the observation. This new quantum state $|\beta,t_0\rangle$ is understandable for the observer because of observation, but it is not the initial state $|\alpha,t_0\rangle$! It is the second challenge of QFT.

This inconsistency in quantum theory reminds the reader of Godel's incompleteness theorem\footnote{Existence of the inherent limitations for every axiomatic system of the arithmetic models.}. This inconsistency can be properly resolved for laboratory systems by choosing the quantum state $|\beta,t_0\rangle$ as the new initial state. Under the new interaction $H_n$ on the quantum states $|\beta,t_0\rangle$, due to the unitarity principle $(U U^{\dag}=1=U^{\dag}U)$, one can obtain the time evolution of this quantum states with certainty: $$\vert\beta,t\rangle=U(t,t_0; H_n)\vert\beta,t_0\rangle\,.$$
The critical point in quantum engineering or technology lies here. For the construction of the technology, we need the deterministic phenomenon, which already exists, $\vert\beta,t\rangle$ \cite{takook18}. We call this the quantum perception against the classical perception, which will be discussed in section \ref{section6}. This inconsistency of observation in quantum theory is quantum gravity's most significant conceptual challenge.

In QFT, however, the perturbation caused by the observer cannot be neglected. An isolated physical system is described by a quantum state, which is a superposition of the Hilbert space bases. Upon observation, the quantum state collapses into the new superposition states. In quantum theory, systems are always assumed to be isolated; however, in the presence of a gravitational field, the system is no longer isolated. The system cannot be shielded from gravity. Therefore, gravity causes to collapse of the quantum state, and the quantum effects become intractable. In the presence of gravity, the physical system, observer, and the gravitational field are all considered as a single isolated system. Again, this new system is in a superposition of states where the observer and the initial system were intertwined, which is not understandable to the observer.

Let us consider a simple instance; for simplicity, we assume that the Hamiltonian is time-independent. If the initial state is one of the Hamiltonian eigenstates $\vert\beta,t_0\rangle=\vert m\rangle$ ($H\vert m\rangle=E_m \vert m\rangle$) then in the time evolution, the quantum state of the physical system is called a ''single state'': $\vert\beta,t\rangle_p= e^{-i E_m (t-t_0)/\hbar}\vert m\rangle ,$
whereas if the initial state is a superposition of different Hamiltonian eigenstates, $\vert\beta,t_0\rangle=\sum_n c_n\vert n\rangle$, then in the time evolution, the quantum state of the physical system is a complicated superposition state:
$\vert\beta,t\rangle_s= \sum_n c_n e^{-i E_n (t-t_0)/\hbar}\vert n\rangle $.
A measurement of the physical system can be interpreted as an interaction with the physical system. Then, one can define a new total system which includes the measuring apparatus. We do this by introducing the perturbed Hamiltonian $H_T=H+H_i$. If $[H,H_i]=0$ and the initial state is single, then a measurement of the physical system results in a quantum state of the physical system, which is also a single state. In all other cases, measurement results in a quantum state of the physical system, which would be a superposition of states. The single and superposition states from the quantum field theoretical point of view may be interpreted as particles and waves, respectively. However, the content of the superposition states is much broader.

In QFT, the free field Hamiltonian commutes with the number operator. If the initial state becomes one of the eigenstates of the number operator, then the quantum state of the physical system is also the eigenstates of the number operator. So it contains a definite number of particles or quanta. However, suppose the initial state is a superposition of the eigenstates of the particle number operator. In that case, the quantum state of the physical system is a superposition of Hilbert's space bases, and the particle picture is inappropriate. Therefore in QFT, observing what we usually call particles or waves depends on the initial state $\vert\alpha,t_0\rangle$ and the Hamiltonian of the interaction between the detector and the system. Thus we see that the particle-wave duality conundrum, well-known in ordinary quantum mechanics, disappears in the framework of QFT as the physical system is described by a quantum state $\vert\alpha,t\rangle$. Given the Hamiltonian, the initial state $\vert \alpha,t_0 \rangle$, and the axioms of quantum mechanics, the quantum state at any time is entirely determined $\vert \alpha,t \rangle=U(t,t_0;H)\vert \alpha,t_0 \rangle$; however this state can describe all that an observer will measure probabilistically {\it i.e.} outcome of the measurements.

The entanglement between the observer, the physical system, and the measuring apparatus shatters classical physics' naive realism and locality. In the QFT model, the truth of a phenomenon is a quantum state $\vert\alpha,t\rangle$. The observer constructs the reality of a phenomenon. The observed reality may be the particles, classical waves, coherent states, squeezed states, etc. Quantum entanglement poses a severe challenge to our conventional way of thinking and understanding of a phenomenon. It is the fundamental basis for quantum engineering technology such as quantum computation, quantum teleportation, telecommunication, quantum cryptography, and quantum tomography \cite{takook18}.

Local symmetries, or the gauge theory, which has been so popular in the present theoretical physics, facilitates an elegant formulation of the fundamental interaction between different matter-fields and field forces classically by using geometry, group theory, and algebra \cite{weinb,ilio,takook1,ta14}. Three out of four fundamental forces of nature ({\it i.e.} electromagnetic, weak, and strong nuclear forces) have been unified within a gauge theory formulation \cite{weinb}. It is hoped that gauge gravity will make it possible to achieve the complete unification in the super-gravity in the de Sitter universe, which will be discussed in the following sections \cite{rata19,taksup}.

One may generalize the ideas of local symmetry and apply them in the case of Hilbert or Fock spaces. The symmetries in the Hilbert space which are reflected in the unitary irreducible representations (UIR) are the global symmetries; their localization of them allows us to derive quantum interaction fields, which are the known quantum "forces" such as those arising from the Pauli Exclusion Principle and the Casimir Effect. The quantum forces may be interpreted as the curvature in Hilbert or Fock spaces. UIR relates a different set of basis vectors of Hilbert space. It is similar to the coordinate transformation in space-time. However, we must first find the relevant UIR of space-time which has been done (for Minkowski space see \cite{weinb}, and de Sitter space see \cite{takthes,ta14}). This UIR on Hilbert space geometry plays a similar role to the Poincar\'e group or de Sitter group on space-time geometry. Localizing the UIR, identical to the general coordinate transformations, the quantum forces may be extracted from connection coefficients in these cases \cite{tak20}. The differential manifold with infinite dimension is elegantly discussed in \cite{choq}, which is this domain's starting point of research. This analogy between the curved space-time and the curved Hilbert space helps us better understand quantum gravity.

\section{Quantum gravity}\label{qg}

Under the principles of quantum theory, an operator is assigned to a physical quantity. However, an operator cannot be attributed to a point. As well as, the definition of an operator at a point is meaningless. It is a mathematical problem faced at the beginning of the QFT, the field operator cannot be appropriately defined as the state is not normalizable \cite{strweit}. The field is a function of the space-time point, and in the field quantization, an operator cannot be defined precisely, and it is ambiguous from a mathematical point of view. Streater-Wightman defined the field operator $\Phi$ by using a spatio-temporal distribution function in a range or an interval on space-time \cite{strweit}:
\begin{equation}\label{fod}
\Phi(f)=\int_{M_4} d^{4}X f(X)\Phi (X),
\end{equation}
where $f$ is the distribution function in space-time, which is called the test function. In this case, the classical field is a function immersed in space-time, and the quantum field operator is a generalization of the concept of function, {\it i.e.} generalized function \cite{gelshi}.

In quantum gravity or the quantization of space-time geometry, the $4$-dimensional space-time manifold, $M_4$, is a physical entity and should be displayed by an operator. Since an infinite set of points makes space-time, the operators must show these points. Nevertheless, this action is not defined mathematically. Like the field operator, a distribution function for a point must be defined, and the classical concept of a point $ P_M $ on manifold $M_4$ must be replaced with a new quantum concept. Similar to the equation (\ref{fod}), the operator associated to the point $ P_M $ of manifold $ M_4 $ may be defined by helping a distribution function as follows:
\begin{equation} \label{opa}
P_M(f) \equiv \int_{A_n} d\mu(x) f(x)P_M(x),
\end{equation}
where $A_n$ is a flat n-dimensional ambient space where the space-time manifold $M_4$ is immersed in it. Classically the curved manifold $M_4$ can be considered as a $4$-dimensional hyper-surface that immerses in the flat manifold $A_n$. By a well choice of distribution function $f(x)$ and the suitable volume element (measure) $ d\mu(x) $ in $n$-dimensional ambient space $A_n$, this quantum geometry (or gravity) can be identified to the one of the different models which exist.

One of the difficulties in understanding the quantum geometry or geometrical section of the gravitational field is the definition of time. From the classical point of view, time in the physical system is moving towards the future. Its meaning depends on the spatial point and the mass and energy in the universe. It depends on the metrics or observer. However, the quantum state does not depend on the time parameter in quantum geometry since it is a sum or integral over all possible time. Therefore, there are many space-time manifolds $[M_4]_i, (i=1,2,...)$, and the concept of Many-Worlds appears.

\subsection{Many-Worlds Interpretation}

In quantum geometry, a point on $4$-dimensional classical space-time, hyper-surface $M_4$, is not localized on hyper-surface $M_4$. We have a distribution of points in the ambient space $A_n$ with a definite probability distribution function. One direction of time has no meaning, similar to the perspective of a particle in quantum mechanics, which is not localized in a spot. That means there are many hyper-surface $[M_4]_i$ with a probability distribution function where the classical hyper-surface has approximately the maximum probability value. In quantum geometry from the Feynman path integral, we know that the probability distribution between two different configurations of space-time, $[M_4]_a$ and $[M_4]_b$, is a sum or integral over all possible hyper-surface $[M_4]_i$ in the ambient space, this is maybe interpreted as the many-worlds.

We suppose that we have found a classical action or Lagrangian of the universe such as: $S_{t}=S_{g}+S_{ng}+S_{in},$ where $S_{g}$ is the gravitational-geometry section, $S_{ng}$ is the matter-radiation fields or non-geometrical section and $S_{in}$ is the interaction between these two-part. The equation of motion of the universe is obtained from the least action principle $\delta_g S_{t}=0$. If we have the boundary conditions of the universe (initial conditions) and by solving the equation of motion, one can predict classically the future of the universe or the structure of the space-time hyper-surface $M_4$:
\begin{equation} \label{clacu}
S_{t}=S_{g}+S_{ng}+S_{in},\;\; \delta_g S_{t}=0, \mbox{ boundary conditions} \Longrightarrow \mbox{ classical space-time} \;M_4.
\end{equation}

The $4$-dimensional hyper-surface $M_4$ must be quantized for quantum geometry without metric selection. Looking at the quantum geometry from the Feynman path integral, we must put together all four-dimensional hyper-surface in a higher-dimensional ambient space $A_n$. This quantity is the partition function or the transition probability amplitude between two different hyper-surfaces:
\begin{equation}
\langle U_a\vert U_b \rangle =\int _{a}^{b} D\Phi_{g}D\Phi_{ng}e^{iS_{t}/\hbar}.
\end{equation}
We live in a particular world that we do not fully understand classically since the initial conditions are not clear to us, and then it is a difficult task for us to obtain the two hyper-surfaces $[M_{4}]_a$ and $[M_{4}]_b$. Quantum theory teaches us that in the evolution of the universe between two configurations $[M_{4}]_a$ and $[M_{4}]_b$, there exists an infinity of paths or configurations, which may be interpreted as the many worlds. However, this evolution is not in the time direction; therefore, it is hard for a specific observer to understand the universe. As long as the observer is a part of the world, the quantum universe cannot be understandable for him due to the entanglement and the quantum state collapse; for an observer who looks at the universe and is not a part of it, the universe can be roughly examined!

In this case, it is essential to be careful that different metrics can be selected on a hyper-surface that all are equivalent, which is a diffeomorphism invariant. In quantum geometry, the quantum state of the universe is a sum over all hypersurfaces with the nonequivalent metric. Each hyper-surface is classically one universe, and the sum over all hyper-surfaces is a "summation" of the parallel classical universes.

\subsection{Universe quantum state}

The quantum state of the universe or geometry is not a function of space and time. It is instead a function of all possible configurations of all matter-radiation fields, which classically constructed the hyper-surface space-time (\ref{clacu}). Therefore, the fundamental question is: how are the dynamics of the quantum state of the universe being defined? Wheeler-de Witt equation is an attempt to round off this problem \cite{dewitt}. One can obtain the "Hamiltonian" of the universe ($H_u$) from the equation (\ref{clacu}) and solve the equation $H_u \vert U \rangle=0$, which means the state with the minimum energy.
Nevertheless, it is essential to note that the concept of energy is unclear and cannot be well defined in curved space-time since it is observer-dependent in general relativity. In Minkowski space-time, energy has a close relationship with the time definition from the time-like Killing vector field, $E \propto \frac{\partial}{\partial t}$, this relation in quantum mechanics results in uncertainty relation $\Delta E \Delta t \approx \hbar$. The time-like Killing vector field does not exist in curved space-time, which is a significant challenge in quantum field theory in curved space-time for defining the vacuum state. In quantum geometry, the observer is a part of the system; therefore, they are entangled with the world.

If we assume that we have obtained the universe Hilbert (or Fock) space, then the quantum state of the universe can be written as a complex superposition of the entire Hilbert space basis:
\begin{equation}
\vert U \rangle=\sum_{n} c_{n}(U)\vert n \rangle,\;\;\;\;\; 1=\sum_{n} \vert n \rangle \langle n\vert,
\end{equation}
where $ \vert n \rangle $ is a complete basis of the universe Hilbert space, a set of discrete and continuous quantum numbers. The sum symbolizes integration for the continuous quantum number and the usual sum for the discrete range. It is difficult to find the coefficients $c_{n}(U)$ or the universe quantum state $\vert U \rangle$ since the boundary conditions of the universe quantum state are obscure.

In other words, the quantum state $\vert U \rangle$ is incomprehensible to an observer because the observer is part of the universe and, as such, is entangled with it, and no sooner has the observer observed the universe than it collapses into a new quantum state, say $\vert u,t \rangle$, where time $t$ is defined by the observer. This new state may be understandable for the observer, but the quantum state of the whole universe before observation $(\vert U \rangle)$ is not. On the other hand, by observing the universe from an observer, the quantum state of the universe $\vert U \rangle$ collapse to a new quantum state $\vert u,t \rangle$ and the primary state $\vert U \rangle$ is not available for the observer:
$$ |U\rangle + \mbox{observations}\;\;\; \Longrightarrow\;\;\;|u,t\rangle\, . $$

The observations of different observers are their actualities, and none of them can be said to represent the truth $(\vert U \rangle)$. The quantum state of the universe is a complex superposition of Hilbert's space bases, which is beyond the reach of any observer who can only observe the reduced form of the quantum state. Moreover, no observer is, in this respect, preferred to another since the quantum state collapse upon the act of observation is an entirely general feature of quantum theory.

At this point, we would like to remark concerning the initial state of the universe: To have the state of the universe at any time (even in principle), we need to know the universe's initial state. However, to get the initial state, it has to be observed, but through the observation of the initial state, it collapses into the new quantum state. It is, perhaps, the most significant conceptual challenge of quantum gravity.

Conceptually, the most significant difficulty stems from the fact that the observer and the system are intertwined. Once the observer is picked, the initial system collapses into another state. The workers in this field try to avoid this insoluble problem by partially splitting the metric into a classical part and a quantum part, the background field method, in the spirit of the quantum fields theory in curved space-time. Therefore, we shall also need to discuss the fundamental concepts of linear quantum gravity in curved space-time, {\it i.e.} vacuum state and infrared divergence, which will be addressed in the following subsection.

\subsection{Linear quantum gravity}

From the usual point of view, if we want to construct the quantum gravity, then the metric $g_{\mu\nu}$ must be quantized. Problems start from here, and quantum gravity fails. The metric classically defines the light cone, and the causality principle, {\it i.e.} sending a message between two points in space-time is less than the speed of light. In QFT, the commutation relation or operator algebra between the quantum field operator and its generalized momentum is defined by space-like separation or causality principle. By using the operator algebra, Hilbert space is constructed. With the metric quantization, we do not have an exact definition of the light cone and the causality principle, whether the light cone is fluctuating or a superposition of numerous states. Thus, the space-like separated points can not be defined precisely, and the operator algebra cannot be determined.

To overcome the mentioned problems, de Witt suggested that, similar to the investigation of the hydrogen atom, use the background field method. The metric decomposes into two parts; the classical background metric and quantum gravitational waves: $g_{\mu\nu}=g^{BG}_{\mu\nu}+h_{\mu\nu}$.
The light cone and the causality principle are defined according to the classical background metric $g^{BG}_{\mu\nu} $ and $h_{\mu\nu} $ is the quantum gravitational waves. In this semi-classical method, geometry, which is characterized by a metric $g^{BG}_{\mu\nu} $, is classical, and the gravitational waves $ h_{\mu\nu} $ are similar to the matter-radiation fields and must be quantized.

For the three known forces in nature, ultraviolet and infrared divergences can be eliminated from the theory and find a renormalizable approach that matches the experimental data. However, in the quantization of the gravitational waves $ h_{\mu\nu} $, whose equation of motion is obtained from linear Einstein's equations, one cannot remove infinity, and the theory of quantum gravity is not renormalizable. Until the first approximation of $ \hbar $ (one loop), it is possible to remove ultraviolet divergence. To not violate the principle of covariance until this approximation, the background metric must be curved. Then we must consider the quantum field theory in curved space-time.

A time-like Killing vector field does not exist in a general curved space-time, and the concept of energy cannot be defined globally. Therefore the vacuum state and time evolution are observer-dependent and cannot define globally. The idea of particles and their number is also observer-dependent. The vacuum state in Minkowski space is defined globally on the hyper-surface of space-time, but it is impossible to define a vacuum state globally in curved space-time.

The adiabatic vacuum state in curved space-time is defined locally and can be manipulated by the ultraviolet divergence in curved space \cite{bida}. The infrared singularity cannot be considered in curved space-time with the adiabatic vacuum state since this vacuum state is defined locally and can be used for probing the local behavior of space-time. It may be possible to bypass the ultraviolet divergence in curved space by redefining the action or Lagrangian for the gravitational field (such as $f(R)$ gravity, super-conformal gravity, or another theory). However, it is impossible to eliminate the infrared divergence with the local vacuum state. This problem may be solved in the de Sitter universe since a global vacuum state exists due to the maximum symmetry of space-time, which will be discussed in the following section.
The experimental cosmological data also confirm that our universe in the first approximation is a de Sitter space-time. Then in the following section, we only discuss the conceptual properties of the QFT in the de Sitter universe.

\section{ QFT in de Sitter space}

de Sitter (dS) space is a curved space-time with the constant curvature ($R=12H^2$) which is the vacuum solution ($T_{\mu\nu}=0$) of the Einstein equation with the positive cosmological constant (\ref{eeq}). Here $H$ is Hubble constant parameter. The dS space-time can be identified with a $4$-dimensional hyperboloid embedded in the $5$-dimensional Minkowski space-time:
\begin{equation} \label{dSs} M_H=\{x \in \R^5| \; \; x \cdot x=\eta_{\alpha\beta} x^\alpha
x^\beta =-H^{-2}\equiv x^2\},\;\; \alpha,\beta=0,1,2,3,4, \end{equation}
where $\eta_{\alpha\beta}=$diag$(1,-1,-1,-1,-1)$. The dS metric is
\begin{equation} \label{dsmet} ds^2=\eta_{\alpha\beta}dx^{\alpha}dx^{\beta}|_{x^2=-H^{-2}}=
g_{\mu\nu}^{dS}dX^{\mu}dX^{\nu},\;\; \mu=0,1,2,3, \end{equation}
where $X^\mu$ is the set of $4$-space-time intrinsic coordinates on the dS hyperboloid. $x^\alpha$ is the $5$-dimensional Minkowski space-time with the condition $x\cdot x=-H^{-2}$, which together constitute the ambient space formalism. For better understanding, one can imagine a shell of a two-dimensional sphere. On the shell one can choose the coordinates systems $(\theta, \phi)$ (intrinsic coordinates) or $(x,y,z)$ with the condition $x^2+y^2+z^2=r^2$ (ambient space formalism). The physics of the shell is independent of the choice of these two formalisms. The quantum geometry of the $2$-sphere can be considered as a sum or integral over all 2-sphere with different radii in three-dimensional flat space.

In dS space-time, a global time-like Killing vector field does not exist, so the equation (\ref{shoreq}) cannot be used to obtain the time evolution of the quantum state in this case. Can one define a unique vacuum state? The authors usually use three types of vacuum states. 1) Bunch-Device or Hawking-Elisse vacuum state is used for massive fields, the massless conformally coupled scalar field, massless spinor field, and photons \cite{bida,allen85,alfo87,chta68,brgamo94,brmo96}. For the massless minimally coupled scalar field, massless spin-$\frac{3}{2}$ field
and linear gravity, two vacuum states are usually used: 2) $O(4)$-invariant vacuum state \cite{allen85,alfo87}, and 3) Gupta-Bleuler or Krein vacuum state \cite{dere98,gareta00,gasiyou10}. It is clear that the second one breaks the dS invariant, and the infrared divergence appears. The third one preserves the dS invariant, and infrared divergence does not also appear \cite{ta14,taklin,taro12}. The Bunch-Device vacuum state can also be used in ambient space formalism \cite{taro15}. The analyticity properties are very crucial for considering the interaction quantum field calculations.

The massless spin-$\frac{3}{2}$ field and linear gravity can be written in terms of the massless minimally coupled scalar field \cite{taklin,taro12,taro15,paenta16,fatata14}. The maximally symmetric space property and the ambient space formalism permit us to write the massless minimally coupled scalar field ($\Phi_{\mathrm{m}}$) in terms of a massless conformally coupled scalar field ($\Phi_{\mathrm{c}}$) by using the following magic relation \cite{ta14,fatata14}:
\begin{equation}
\label{mmcmcc}
\Phi_{\mathrm{m}}(x)\equiv \left[Z\cdot\partial^\top + 2H^2 Z\cdot x\right]\Phi_{\mathrm{c}}(x)\, ,
\end{equation}
where $Z^\alpha$ is an arbitrary constant five-vector, and $\partial^\top_\alpha$ is a tangential derivative on the dS hyperboloid. Therefore a unique vacuum state can be defined for all of the quantum fields in dS space {\it i.e.} Bunch-Device vacuum state \cite{taro15}. But in this vacuum state, the theory is not renormalizable, and only in the Kerin vacuum state can a renormalizable theory be constructed.

The QFT in de Sitter space in ambient space formalism is straightforward and similar to the Minkowski counterpart. This formalism allows us to reformulate QFT in a rigorous mathematical framework, based on the analyticity of the complex pseudo-Riemannian manifold and the group representation theory, similar to the QFT in the Minkowski space-time \cite{ta14}. The one-particle Hilbert space can be globally constructed from the UIR of the de Sitter group. Then the vacuum state can be defined globally and uniquely \cite{ta14}. Therefore, it is now possible to bypass the infrared divergence in the ambient space formalism since the vacuum state is defined globally \cite{taklin,taro12,taro15,pejrahb16}. The infrared divergence is due to the new "gauge-invariant" of the minimally coupled scalar field. This new symmetry, likewise local symmetries, represents the interaction between scalar and spinor fields in this formalism \cite{gata18}.

In this space-time, the super-gravity cannot be formulated without the vector-ghosts field in the intrinsic coordinate system \cite{pilch} {\it i.e.} a real classical Lagrangian does not exist. But this problem is also solved in ambient space formalism \cite{taksup,paenta16,tasual}, since the spinor and its charge conjugation are defined directly from the dS universal covering group, $Sp(2,2)$, \cite{tasual,tacha,bagamota01}. In the theory of super-gravity in the de Sitter space-time, the massless spin-2, spin-3/2, and minimally coupled scalar fields play a crucial role and are treated as gauge fields. These three fields are inextricably linked together and appear in a complete theory of quantum gravity based on the gauge theory \cite{ta14,paenta16,rota05}.

The next problem which can be easily manipulated in ambient space formalism is the conformal invariant or conformal gravity. For the conformal invariant, the linear part of the gravitational field must be represented by a rank-3 mixed symmetry tensor field \cite{bfh83,amta19}. From the dS gauge gravity in ambient space formalism, a rank-3 mixed symmetry tensor field can be constructed as a gauge potential \cite{rata19,ta14} then the conformal invariant may be conserved \cite{fatata14,tata,gagarota08,berotata}. The other problems which can be solved in this formalism are: a finite total entropy of quantum fields in dS space \cite{taen}, quantum dS-black hole \cite{takdsbl,taro12,taro15} and $N=4$ dS conformal super-gravity \cite{rata19,ta14,kalosh}.

It is important to note that although this formalism may solve some problems of quantum gravity, there is always the challenge of quantum state collapse and the definition of the initial conditions in the presence of gravity for an observer. In the next section, the conceptual problem of perception for an observer in curved spacetime will be discussed.

\section{Perception in classical and quantum theory} \label{section6}

As an observer of the natural world, what are our limitations in understanding the physical system? What is the meaning of understanding and perception of a phenomenon? And how do these differ in the quantum and classical senses? One of man's perennial concerns has been knowing and understanding the world outside the mind. This ambition is not, in all likelihood, going to change in the near or distant future. It may well be that the root of all wars, bloodshed, erection of citadels of unjust rulers, and many similar afflictions was inadequate understanding of the natural phenomenon, assisted by base human traits such as greed and extreme selfishness.

Moreover, now that humankind seems to have reached a certain kind of maturity and perhaps the highest degree of scientific and technological advancement, the question arises: Have we got to the point where we know that understanding the truth of a natural phenomenon is inaccessible to humans? Does anyone understand that its reality is not unique? And therefore, it is the end of all wars and violence achievable. What exactly do we mean when we say we know something? There are two kinds of understanding of a phenomenon: classical and quantum theory understandings.

Let us first sum up the classical case. According Avicenna\footnote{Avicenna (980-1037) was a Persian polymath who is regarded as one of the most significant physicians, astronomers, thinkers, and writers. He has been described as the father of early modern medicine.} perception of a thing consists of the encapsulation of the form of that thing in the mind, implying that perception can only come after observation. In a more technical sense, understanding explains the system's behavior in terms of the theory's postulates by a mathematical model. In other words, we must have a logical connection between the observations and the explanation of the system's behavior, an axiomatic formulation. A lack of logical connection will mean we must seek another axiomatic formulation. In the classical view, the reality of a phenomenon is unique, and the act of observation only reveals or discovers it.

Regarding our perception of reality, we note that perception follows from observation in the classical physics paradigm. The classical theory seeks to describe what is observed and predict what is to come, all based on classical mechanics. Thus a consistent description of the physical world or all observations constitutes an understanding of a phenomenon. In this sense, the observer thinks that his reality is precisely the truth and the difference between his reality with the other observer's reality is the origin of the above problems in the first paragraph of this section.

From the quantum theory point of view, the concept of understanding is such a highly sophisticated process that we must either abandon it or else redefine it. Classical understanding follows from observing natural objects, but according to quantum mechanics, reality cannot be said to exist before the act of measurement, and the latter changes the system's quantum state. It is the act of observation by an observer that makes reality. According to the principles of quantum mechanics, the physical system is described by the state ket $| \alpha \rangle $, which for an isolated system is a superposition of the relevant Hilbert or Fock space basis vectors, $|\alpha\rangle =\sum_n |n\rangle \langle n|\alpha\rangle $. In the quantum theory paradigm, understanding must be defined based upon the quantum state $\vert \alpha \rangle$; this quantum state allows a probabilistic description of all observations or different possible realities in the most elegant manner.

Quantum theory teaches us that while the facts and observations are observer-dependent, the quantum state $\vert \alpha \rangle$ is unique or observer-independent. It is very similar to the hypersurface manifold of space-time, which is unique or observer-independent. Still, metrics on it are observer-dependent. The quantum state is a superposition of the basis vectors in the Hilbert or Fock spaces; it is not comprehensible in the classical mechanical paradigm because the act of observation causes the quantum state to "collapse" (or "reduce") into one of the superposed states $\vert \beta,t_0\rangle$. The latter form is classically understandable, but the state of the system before measurement, $| \alpha\rangle $, (and hence collapse) is not.

The superposition of states implies that the system is at all the quantum states' basis but with different intensities. The conventional mind of modern man finds it very difficult to accept the concept of "everywhere or several states at once" and finds the idea of superposition unintelligible. Since the quantum state $| \alpha, t \rangle $ can probabilistically explain all our possible observations, the quantum state $| \alpha, t \rangle $ may be understandable after various statements and obtaining the probability of each quantum state basis.

Can we understand the quantum state $| \alpha, t \rangle$? Classically, the answer is a resounding "no," because, as previously mentioned, understanding a phenomenon in the classical sense can only come after observation. Still, the quantum state collapses as soon as an observation is made. However, the answer to the above question is affirmative if we have the initial state. The system is in a superposition of all the states, and the probability for each of these states is $|c_n|^2$. Moreover, all these states are interrelated since $\sum_n |c_n|^2= 1$. It implies that a variation of the others necessarily accompanies any of these coefficients; this is also related to the question of quantum entanglement, which defies classical understanding. Quantum entanglement has been observed in countless situations and is a confirmed fact (see \cite{takook18}). This entanglement (a superposition of product states) is the greatest challenge to the conventional way our minds have grown accustomed to thinking and understanding things.

Thus "classical understanding" stems straight from observation and is mathematically expressed in the Lagrangian system, whereas "quantum understanding" is attained through the quantum state $| \alpha, t \rangle$. We might further assert that "classical understanding" is based on observed actualities, whereas "quantum understanding" is based on quantum states. It is these states which express the true nature of things and that there are no "actual objects" before the measurement; the actualization occurs with the act of measurement, whereas the quantum state $| \alpha, t \rangle$, which encapsulates the true nature of the system, exists before the act of measurement and is altered by it. The quantum state $| \alpha, t \rangle$ explains every actuality that we know about the physical object ({\it i.e.} the system) in as much as it is a superposition of numerous states. To the conventionally trained mind, the concepts of "superposition of states" and "entanglement" are problematic for the human mind. At any rate, we stress that the critical point at issue here is a kind of "absolute truth" represented by the quantum state, which is partly lost to us every time we try to encompass it. Further, we can assert that the actualities in the physical world are none other than collapsed quantum states.

The limitations of the human mind are numerous. As a pertinent example, consider a globe with a curved two-dimensional surface. Our minds can envisage this object and understand it, but our minds have difficulty envisaging and understanding a three-dimensional sphere (in $4$D), even though such "sphere" properties are entirely known. There is no imagination of such a "sphere" in mind even though it is completely familiar to, and well known by, mathematicians.
In Minkowski space and experimental physical systems, where the initial state can be constructed, the quantum state $|\alpha, t \rangle$ can be understood. This state is essential to quantum engineering \cite{takook18}. However, the problem is serious for the quantum universe since the initial state does not in our ability to construct.

Let the universe consider an isolated system whose quantum state is $\vert U \rangle$. Any observer cannot observe this quantum state, and, in particular, it cannot be observed by those observers who are (necessarily) a part of the isolated system. $\vert U \rangle$ can not be a function of space and time because it is summed over all the possible space and time variables. It is, therefore, only dependent upon the configuration of all the different fields of which the universe is composed. However, as soon as a metric tensor is specified, the universe's quantum state collapses into a new state. The new state is the universe's quantum state for an observer who perceives it. The metric or the observer can be treated as a background field, which will not detract from the effectiveness of quantum field theory in curved space.  Because the quantum state collapse occurs only concerning the Hamiltonian of the measuring apparatus, which is treated as part of the background. If at a given time $t_0$, the quantum state $|\alpha, t_0 \rangle$ be known, the quantum state at any other time $t$, $|\alpha, t\rangle$, can be found. Thus everything the cosmologists do may be interpreted as a quest for the initial quantum state, $|\alpha, t_0\rangle$, which is a complicated task!

\section{Conclusion}

Although the technical problems of quantum gravity may be solved and a formulation of it may be found, the goal of an ultimate understanding of the universe or conceptual problems cannot be accomplished by the quantum field theory. The quantum state of the universe $|U \rangle$ is not understandable for an observer due to the entanglement of the observer and gravity, collapsing the quantum state by observation and a lack of the universe's initial state. Conceptual problems of quantum gravity may be only understandable in the approximation method for an observer far from a black hole and or in the laboratory system. In quantum theory, it is impossible to reach the certainty for explication of reality. However, it is still the most qualified referee to address the problem, thereby saving man from illusion, particularly the illusion of certainty and the ultimate truth.

The classical understanding is achieved by knowing the reality. Still, the quantum mechanical understanding is achieved by knowing the quantum state $|\alpha, t \rangle$, with all its limitations, which can be explicitly accomplished for the laboratory system. The quest to understand our relationship with the rest of the world would always be achieved by unification with the rest.

The gravitational field has no inconsistency with quantum mechanics principally, and quantum gravity is not visible to us except in the approximate situation since we are also a part of the source of the gravitational field. The quantum theory has inconsistent principles {\it i.e.} relation between initial state and collapse of quantum state due to the observation, which becomes a severe challenge for quantum gravity.

\vspace{0.5cm}

{\bf{Acknowledgements}}: The author wishes to express his particular thanks to M.R. Tanhayi and S. Tehrani-Nasab for helpful discussions.

\end{document}